\documentclass[twocolumn,showpacs,amsmath,amssymb,floatfix]{revtex4}

\usepackage{graphicx}% Include figure files
\usepackage{dcolumn}% Align table columns on decimal point
\usepackage{bm}% bold math

\def\be{\begin{equation}}
\def\ee{\end{equation}}
\def\bes{\begin{equation*}}
\def\ees{\end{equation*}}
\def\bea{\begin{eqnarray}}
\def\eea{\end{eqnarray}}
\def\beas{\begin{eqnarray*}}
\def\eeas{\end{eqnarray*}}

\begin{document}

\title{Theory of NMR in semiconductor quantum point contact devices}

\author{N. R. Cooper$^{1}$ and V. Tripathi$^2$}

\affiliation{$^1$T.C.M. Group, Cavendish
Laboratory, J. J. Thomson Avenue, Cambridge CB3 0HE, United Kingdom.\\
$^2$ Department of Theoretical Physics, Tata Institute of Fundamental
Research, Homi Bhabha Road, Mumbai 400005, India.}

\date{October 23, 2007}

\begin{abstract}

We describe how a local non-equilibrium nuclear polarisation can be
generated and detected by electrical means in a semiconductor quantum
point contact device. We show that measurements of the nuclear spin
relaxation rate will provide clear signatures of the interaction
mechanism underlying the ``0.7'' conductance anomaly.
Our analysis illustrates how nuclear magnetic
resonance methods, which are used extensively to study
strongly-correlated electron phases in bulk materials, can be made to
play a similarly important role in nanoscale devices.

\end{abstract}

\pacs{73.21.Hb, 73.21.-b,  76.60.-k, 85.30.Hi}

% 73.21.Hb 	Quantum wires
%
% 73.21.-b Electron states and collective excitations in
%multilayers, quantum wells, mesoscopic, and nanoscale systems (for
%electron states in nanoscale materials, see 73.22.-f)
%
%76.60.-k Nuclear
%magnetic resonance and relaxation (see also 33.25.+k Nuclear resonance
%and relaxation in atomic and molecular physics and 82.56.-b Nuclear
%magnetic resonance in physical chemistry and chemical physics)
%
%73.43.Fj 	Novel experimental methods; measurements
%
%85.30.Hi 	Surface barrier, boundary, and point contact devices

\maketitle

The confinement of electrons to nanoscale regions in semiconductor
devices leads to the formation of low-dimensional quantum systems
which are highly susceptible to quantum fluctuations.
Electron-electron interactions can then have dramatic effects.
Indeed, experimental studies of nanoscale semiconductor devices have
uncovered evidence of many very interesting strong-correlation
phenomena -- including the
Kondo\cite{goldhaber} effect, spin-charge separation\cite{auslaender}
and the ``0.7 effect''\cite{ThomasNSPMR96}.

The 0.7 effect refers to a series of anomalous features that are
observed in the conductance of quantum point contact (QPC)
devices\cite{ThomasNSPMR96,Cronenwett}. The
QPC is a simple device, in which a split gate is used to
confine the electrons in a two-dimensional electron gas into a 
 quasi one-dimensional
(1D) channel.  The anomalous conductance features observed are believed to
arise from electron-electron interactions in this quasi-1D geometry,
but the question of how interactions lead to these conductance
anomalies is hotly
debated.
Many theoretical models have been proposed to account for the
observations\cite{theory,reillyth,Meir,matveev,rejecmeir}: including spontaneous spin
polarisation\cite{reillyexpt,reillyth}, the Kondo
effect\cite{Cronenwett,Meir}, and the spin-incoherent Luttinger
liquid\cite{matveev}. These theories make similar predictions for the
conductance, the property that is usually measured, precluding a
conclusive experimental distinction between them.

In this paper we show that the nature of the electronic state
responsible for the 0.7 effect can be uncovered through a variant of
nuclear magnetic resonance (NMR). NMR is a very powerful tool that is
widely used to study
  strongly-correlated electronic phases in bulk materials.
The very small active
volume in a QPC 
 would make it extremely difficult to perform a
conventional NMR measurement, owing to the small number of nuclei
coupled to the electrons.
Here we describe how NMR can be performed on a QPC by generating
and detecting a {\it local} non-equilibrium nuclear spin polarisation.
We then turn to discuss the nuclear spin relaxation rate in the vicinity of
the 0.7 effect.  We show that different interaction mechanisms that can lead
to similar features in conductance have very different effects on the nuclear
spin relaxation rate.  We identify clear experimental signatures which
distinguish between different proposed scenarios for the 0.7
effect.
Our work shows how electrical manipulation of local nuclear spin
polarisation opens the possibility of performing NMR 
in nanoscale
electronic systems.

The local NMR scheme that we propose relies on the possibility to generate a
non-equilibrium nuclear spin polarisation in the vicinity of a QPC. This can
be achieved in various ways using purely electrical means.  Current-induced
breakdown of the $\nu=2/3$ fractional quantum Hall state\cite{kronmuller1} has
been used to create large local nuclear polarisations in
QPCs\cite{yusa}.  Alternatively, a non-equilibrium nuclear
polarisation can be achieved by the selective backscattering of the
spin-polarised edge states of the $\nu=2$ quantum Hall
state\cite{Wald,Dixon,machida}; this creates a region in which the edge states
are out of spin equilibrium and their relaxation leads to a local dynamic
nuclear spin polarisation. Simple gate geometries can be envisaged for which
this non-equilibrium spin polarisation is placed at the centre of a second
point contact.  In both these methods the dynamic nuclear polarisation is
generated at non-zero magnetic field, so its use in probing electron systems
at low field would require a field sweep that is shorter than the nuclear spin
equilibration time\cite{footnote1}.
A recent proposal has shown that, through
spin-orbit coupling, it is possible to generate a local dynamic nuclear spin
polarisation in a biased quantum wire at zero magnetic field\cite{tripathicc}.

The presence of nuclear polarisation in the vicinity of the QPC can be
detected by its effect on the two-terminal conductance.  The effect
arises from the Overhauser shift of the electronic Zeeman energy from
hyperfine contact interactions.  
To estimate
the sensitivity of this resistive detection scheme for the QPC, we
model the device by a quasi 1D wire with Hamiltonian
\begin{equation}
 H = \sum_{s,k,\sigma} \left[\epsilon_s + \frac{\hbar^2 k^2}{2m}
  +\frac{\sigma}{2}g\mu_B B\right]\hat{c}^\dag_{nk\sigma}\hat{c}_{nk\sigma}
+
A_{\rm s} \sum_i \vec{I}_i\cdot \vec{S}({\bf R}_i) 
\label{eq:ham}
\end{equation}
where $\epsilon_s \equiv \hbar\omega_y(s+1/2)$ are the edges of the in-plane
subbands (we assume that the out-of-plane subband spacing $\hbar\omega_z$ is
very large), $\sigma=\pm$ are the spin polarisations. $\vec{I}_i$ is the spin
of nucleus $i$ at location ${\bf R}_i$, which is coupled to the electronic
spin density $\vec{S}({\bf r})$ via the hyperfine contact interaction.  The
net electron Zeeman energy, with Overhauser shift, is therefore
\begin{equation}
Z_{\rm e} \equiv g\mu_B B + A_{\rm s} n_{\rm
  nuc} \langle I^z\rangle
\label{eq:zeeman}
\end{equation}
where $n_{\rm nuc}$ is the nuclear density.
Throughout this work we neglect the nuclear Zeeman
energy,  assuming it to be small compared to electronic energy
scales. For quantitative estimates we choose parameters for typical GaAs
QPCs\cite{footnote2}.

The linear conductance of the quasi 1D wire is
\begin{equation}
G(\mu,Z) = \frac{e^2}{h} \sum_{s,\sigma} f_{\rm }
\left(\epsilon_s +\sigma Z_{\rm e}/2\right)
\end{equation}
 where $ f_{\rm
  }(z) \equiv [e^{(z-\mu)/k_BT}+1]^{-1}$, $T$ is the temperature and $\mu$ the chemical
potential. 
Even a small change in nuclear polarisation, and hence $Z_{\rm e}$ (\ref{eq:zeeman}), can lead to a sizeable change of
conductance.  
Owing to the importance of electrostatic forces, a fixed gate voltage, $V$, on
the QPC fixes the 1D electron density in the channel, $n = c(V-V_0)/e$ ($c$ is the
capacitance per unit length and $V_0$ the pinch-off voltage).
Thus, the sensitivity of $G$ to small changes in nuclear polarisation 
at fixed gate
voltage can best be
expressed  by the derivative of $G$ with respect to
$Z_{\rm e}$ {\it at fixed particle density, $n$}.

\begin{figure}
\includegraphics[width=8.6cm]{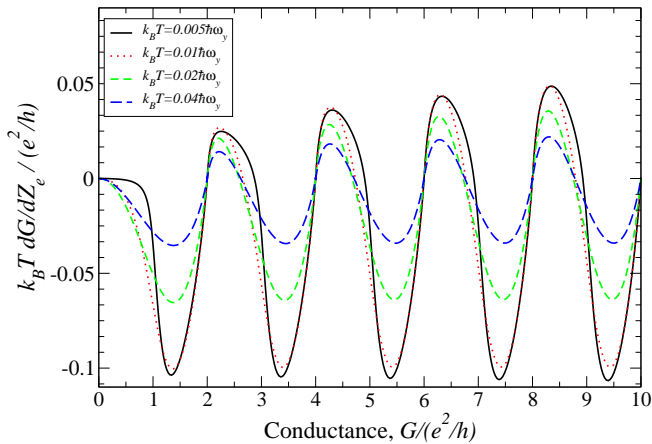}
\caption{\label{fig:modulation} 
The sensitivity of the QPC device to small changes in the nuclear polarisation
at fixed gate voltage is
conveniently represented by $\left.\frac{d G}{d Z_{\rm e}}\right|_n$.
 This is plotted as
a function
the conductance
(which varies with  gate voltage), for a non-interacting
electron gas with subband
spacing
$\hbar\omega_y$ and $Z_{\rm e} = 0.03\hbar\omega_y$.
}

\end{figure}

Fig.~\ref{fig:modulation} shows 
$\left.\frac{\partial G}{\partial
    Z_{\rm e}}\right|_n$
over a range of
temperatures
as a function of $G$ (which 
eliminates device-specific properties). 
 The conductance is most sensitive to
changes in $Z_{\rm e}$ when $G$ is away from a quantised value. The
conductance changes by $\sim 0.01 e^2/h$ 
when $Z_{\rm e}$ changes by $\sim k_BT/10$: for GaAs at $T=50\mbox{mK}$, this corresponds to a
change in the fractional nuclear polarisation of $\Delta I^z/(2I)\simeq 0.3\%$.

\begin{figure*}
\centering
\includegraphics[width=17cm]{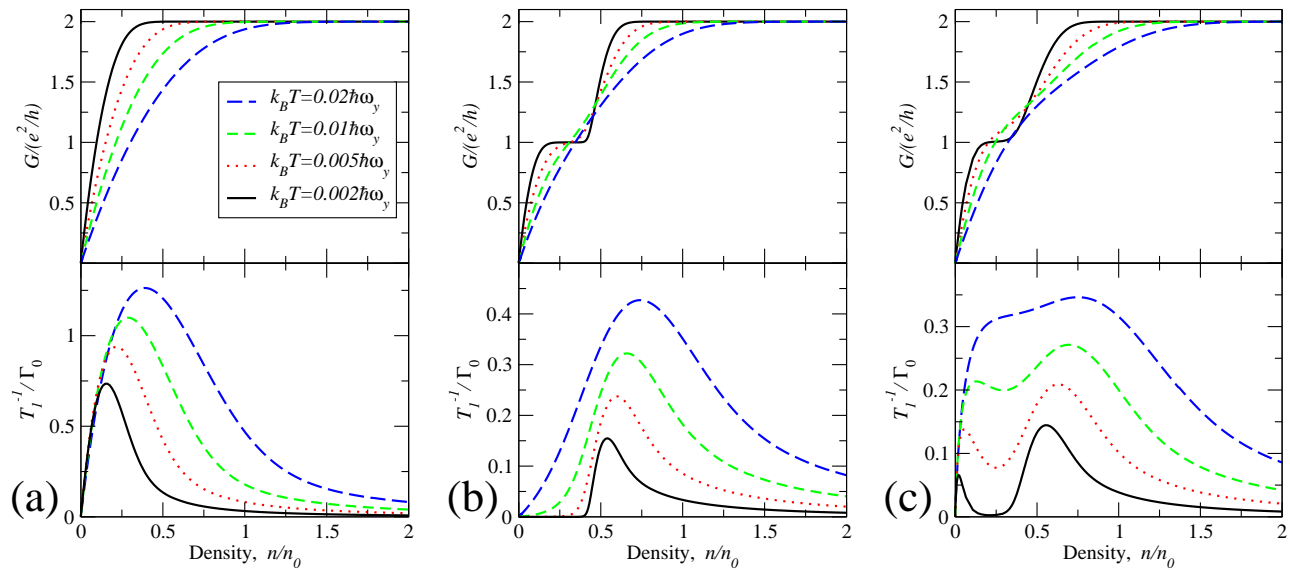}

\caption{\label{fig:relax}
Conductance (top panels) and nuclear spin relaxation rate (bottom panels) for
a quasi-1D electron gas  on the first conductance riser.
(a) Non-interacting electron gas, with small Zeeman
  energy, $Z_{\rm
    e}=0.001 \hbar\omega_y$. 
(b) Non-interacting electron gas, with larger Zeeman
  energy, $Z_{\rm
    e}=0.05 \hbar\omega_y$. 
 (c) Electron gas with exchange-enhanced spin-splitting,
 Eqn.~(\ref{eq:exchange}) with $Z_{\rm e}=0.001 \hbar\omega_y$ and $\gamma
 =0.1 \hbar\omega_y/n_0$. 
The electron density $n$ is in units of $n_0\equiv \sqrt{m\omega_y/\pi h}$.
 A typical quantum wire has subband spacing $\hbar\omega_y =
 20\mbox{K}$\protect\cite{abigraham}, for which the illustrated temperatures
 are $T=40, 100, 200 \mbox{ and } 400\mbox{mK}$.  }

\end{figure*}

By monitoring the rate of change of the two-terminal conductance, the nuclear
spin relaxation rate, $T_1^{-1}$, for nuclei in the QPC may be measured.
Nuclear spin relaxation is dominated by coupling to the electrons, and
is determined by their low-frequency spin dynamics, via\cite{narath}
\begin{equation}
T_1^{-1}({\bf R}) 
=\frac{A_{\rm s}^{2}}{2\hbar^{2}}\int_{-\infty}^{\infty}dt\,
\langle
S^{+}({\bf R},t)S^{-}({\bf R},0)\rangle\label{eq:kubo}
\end{equation}
where the angled brackets denote thermal and quantum averages. Since
there are many nuclei per electron (of order $10^6$) the gradual nuclear
depolarisation leads to a smooth evolution of $G$ over the
timescale $T_1$; this time is much longer than electronic timescales (see below) so many
electrons pass through the QPC and contribute to the measurement of $G$. 

We shall calculate the nuclear spin relaxation rate for an electron gas on the
first conductance riser, $0 \leq G\leq 2e^2/h$\cite{footnote3},
where experiments
show the appearance of the anomalous conductance features of the ``0.7 effect''\cite{ThomasNSPMR96}.  We first calculate the nuclear spin relaxation rate for a
non-interacting electron gas, before turning to consider the effects of
electron-electron interactions within several theoretical models of the 0.7 effect.

{\bf (i) Non-interacting 1D electron gas}  

We consider a non-interacting electron gas described by (\ref{eq:ham}).
Restricting attention to the lowest subband ($s=0$), and focusing on nuclei at
the centre of the quantum wire,
we find
from (\ref{eq:kubo})
\begin{equation}
{T_1}^{-1} = \Gamma_{0}
\int_{|Z_{\rm e}|/2}^\infty
\frac{ f_{\rm
  }\left(\epsilon\right)\left[1-f_{\rm
    }\left(\epsilon\right)\right]}{\sqrt{\epsilon^2 + (Z_{\rm e}/2)^2}}
d\epsilon
\label{eq:gamma0}
\end{equation}
where 
\begin{equation}
\Gamma_0\equiv \frac{2\pi A_{\rm s}^2 m}{\hbar^3 w_y^2w_z^2}
\end{equation}
is a characteristic rate and $w_yw_z$ is the root mean square transverse area of
the lowest subband.  For a typical GaAs QPC, $\Gamma_0\simeq
0.5\mbox{Hz}$.  The value of $\Gamma_0$ is very sensitive to the value
of $w_yw_z$. Our main results, below, concern the dependence
of $T_1^{-1}$ on gate voltage and temperature and are independent of
this overall scale.

In Fig.~\ref{fig:relax}(a) we show the conductance $G$ and relaxation rate
$T_1^{-1}$ as a function of electron density (controlled by gate voltage) 
 for a small Zeeman energy ($Z_{\rm e}\ll k_BT$).
There is a maximum in $T_1^{-1}$ close to the midpoint of the conductance
riser. Increasing the electronic Zeeman
energy to $Z_{\rm e}\sim k_B T$,  Fig.~\ref{fig:relax}(b), leads to the emergence of a plateau at $G=
e^2/h$; this is accompanied by a suppression of the peak in
$T_1^{-1}$.

It is instructive to compare these results with the conventional Korringa
expression for the nuclear spin relaxation rate of a metal, $T_1^{-1} \propto
\rho_\uparrow \rho_\downarrow T$, where $\rho_{\uparrow,\downarrow}$ are the
densities of states for the two spin species at the Fermi level.  The Korringa
expression applies when $k_B T\ll \mu$, which for the quantum wire implies
that $G\simeq 2 e^2/h$.  In this regime, we do find that $T_1^{-1}\propto
(1/n^2) T$, consistent with the Korringa expression; increasing the Zeeman
energy leads to a small {\it increase} in $T_1^{-1}$, consistent with an
expected increase in $\rho_\uparrow \rho_\downarrow$ at fixed 
$n$.
However, the
Korringa expression does not account for the most dramatic signatures in
$T_1^{-1}$. These occur on the conductance riser, $0\lesssim G\lesssim 2 e^2/h$
where $k_BT\sim \mu$.  In this regime, we find that $T_1^{-1}$ increases
more slowly than linear in $T$.  An
increase in the Zeeman energy leads to a dramatic {\it decrease} in the
height of the peak in $T_1^{-1}$; at the same time, the position of the maximum
shifts to lie in the regime $e^2/h \lesssim G \lesssim 2e^2/h$ where both spin
species are occupied [compare Figs.\ref{fig:relax}(a),(b)].

{\bf (ii) Exchange-enhanced spin-splitting}

Refs.~\cite{reillyexpt,reillyth} have provided a phenomenological theory that
can successfully reproduce many features of the 0.7 effect. The electron gas
is assumed to experience a density-dependent exchange splitting, leading to an
effective Zeeman energy
\begin{equation}
Z_{\rm eff} = Z_{\rm e} + \gamma n
\label{eq:exchange}
\end{equation}
where
$\gamma$ is a phenomenological
parameter. (We note that a linear dependence of exchange energy on $n$
is expected for Coulomb interactions, with $\gamma
\propto e^2/\epsilon$.)
Treating the system as a non-interacting gas with this exchange enhanced Zeeman splitting leads to the 
results shown in Fig.~\ref{fig:relax}(c), where
$\gamma$ has been chosen to give conductance features similar to
those of the 0.7 effect\cite{reillyth}.  
Comparing the results for $T_1^{-1}$ with those for non-interacting electrons
at the same bare Zeeman energy, Fig.~\ref{fig:relax}(a), one sees that the
main effect of exchange is a strong suppression of the peak in $T_1^{-1}$.
This is consistent with the result discussed above, that increasing the Zeeman
energy leads to a suppression of $T_1^{-1}$. However, the exchange enhancement
of the Zeeman energy leads to a qualitatively new feature: there are now two
peaks in $T_1^{-1}$ as a function of density.  At very small densities
exchange interactions are negligible and $T_1^{-1}$ rises as for the
non-interacting gas with small Zeeman energy, Fig.\ref{fig:relax}(a); at
higher densities the increase of exchange splitting at first causes a
reduction in $T_1$, leading to a second peak similar to that for a large
Zeeman energy, Fig.\ref{fig:relax}(b).  {\it The observation of a double-peak
  structure in $T_1^{-1}$ as a function of density (gate voltage) is a clear
  signature  of a density dependent exchange-enhanced spin-splitting.}

{\bf (iii) ``Kondo'' model} 

Within the ``Kondo'' model for the 0.7 effect\cite{Cronenwett,Meir}, one of the
electrons is assumed to become trapped in a quasi-bound state, and to behave as
a spin-1/2 ``impurity'' exchange-coupled to the rest of the electron
gas\cite{Meir,rejecmeir}. This exchange coupling, $J_K$, leads to a low energy scale, the Kondo
temperature, $k_B T_{\rm K}\sim \epsilon_F e^{-1/J_{\rm K}\rho(\epsilon_F)}$
[$\rho(\epsilon_F)$ is the density of states at the Fermi level].  For $Z_{\rm e} \ll k_B T_{\rm K}$, which is the
regime that we shall consider here, the conductance
for the QPC shows an interesting temperature-dependence,
with a crossover from
$G < 2 e^2/h$  for $T \gtrsim T_{\rm K}$ to
$G\simeq 2e^2/h$ for $T\ll T_{\rm K}$\cite{Cronenwett,Meir}.

This crossover should be accompanied by dramatic
changes in the nuclear spin relaxation rate.
The nuclear spin relaxation  in the QPC is dominated by the
fluctuations of the impurity spin\cite{footnote4}.
The fastest rate is for those nuclei located close to the
impurity, which 
are coupled to the impurity spin
with an energy scale
$A_{\rm d}
\sim A_{\rm s}/(w_x w_y w_z)$ where $w_xw_yw_z$ is the mean volume of the
impurity. 
Relating (\ref{eq:kubo}) to the impurity dynamical susceptibility, 
we can make use of known results in limiting cases. 
For $T\gg T_{\rm K}$, the coupling of the impurity spin to the electron gas is
relatively weak. Using the results of Ref.\onlinecite{gotze} 
we find
\begin{equation}
T_{1}^{-1}  = 2\frac{A_{\rm
    d}^{2}S(S+1)}{3\pi\hbar(k_{B}T)[J_{\rm K}\rho(\epsilon_{F})]^{2}},
\label{eq:kondohigh}
\end{equation}
where  $S=1/2$ for the spin-1/2 impurity\cite{footnote5}.
For $T \ll T_{\rm K}$, the Kondo singlet is well-formed and the system behaves
as a
local Fermi liquid. From Ref.~\cite{shiba}, one then recovers a Korringa law for the nuclear
spin relaxation rate, with
\begin{equation}
  T_1^{-1}
  =\frac{2\pi(k_{B}T)A_{\rm d}^{2}}{\hbar(g_{s}\mu_{B})^{4}}\chi_{\text{imp}}^{2},
\label{eq:kondolow}
\end{equation}
where $\chi_{\rm imp}$ is the static Kondo impurity
susceptibility, 
which is a universal function of $T/T_{\rm K}$ and tends to a constant as
$T\to 0$\cite{hewson}.
The nuclear spin
relaxation rate is a non-monotonic function of $T$, passing through a maximum at $T \sim
T_{\rm K}$ with a maximum rate of order
\begin{equation}
\Gamma_{\rm Kondo}\simeq \frac{A_{\rm d}^2}{\hbar k_B T_{\rm K}} = \frac{A_{\rm s}^2}{\hbar k_B
  T_{\rm K} (w_xw_yw_z)^2}
\label{eq:gammakondo}
\end{equation}
{\it This non-monotonic temperature dependence of $T_1^{-1}$ is characteristic of the
Kondo physics}. It is qualitatively distinct from the
case of non-interacting electrons, or electrons with
exchange-enhanced Zeeman energy, for which $T_1^{-1}$ increases monotonically
with $T$.

{\bf (iv) Spin-incoherent Luttinger Liquid}

Finally, we consider the possibility that the electron system in the QPC
behaves as a strongly interacting 1D wire.  Strong repulsive interactions lead to
pronounced local charge density wave order, and a suppression of the
exchange interaction energy scale $J_{\rm LL}$, with $J_{\rm LL}\ll
\epsilon_F$\cite{matveev}.

At low temperatures, $T\ll J_{\rm LL} \ll \epsilon_F$ the system
should behave as a Luttinger liquid. Nevertheless, since it is coupled
to Fermi liquid leads the conductance is $G = 2e^2/h$, and is
insensitive to the electron-electron interactions\cite{maslovstone}.
Applying the general approach of bosonisation to the spin
susceptibility of the repulsive 1D electron gas leads to the
prediction\cite{luttt1} that as $T\to 0$,
$T_1^{-1}\sim T^{K_\rho}$, with $K_\rho < 1$ for repulsive
interactions.  {\it Thus, the nuclear spin relaxation rate is
sensitive to the formation of a Luttinger liquid.}

As temperature is increased, the 1D electron gas enters the regime of
the ``spin-incoherent'' Luttinger liquid\cite{matveev,fiete}, $ J_{\rm
LL} \ll k_B T\ll \epsilon_F$.  The conductance is then
expected\cite{matveev} to be $G\simeq e^2/h$.  The spin-incoherent
Luttinger liquid is characterised by an enhanced nuclear spin
relaxation rate.  This arises from the existence of low energy
spin-flip excitations, of bandwidth $J_{\rm LL}$, which decouple from
the electronic motion. Treating the spin-flip excitations as a
spin-chain with lattice constant $1/n$ and exchange energy $J_{\rm
LL}$ one finds for $k_BT\gg J_{\rm LL}$
\begin{equation}
{T_1}^{-1} \sim \Gamma_{\rm SILL} \equiv \frac{I A_{\rm s}^2 n^2}{\hbar w_y^2 w_z^2 J_{\rm LL}} 
\label{eq:sill}
\end{equation}
Since $J_{\rm LL}\ll \epsilon_F$, the relaxation rate (\ref{eq:sill}) is
parametrically enhanced as compared to that for the non-interacting electron gas
(\ref{eq:gamma0}).  {\it In the spin-incoherent Luttinger liquid regime $T_1^{-1}$ is
  expected to be large and  weakly temperature-dependent.}

In summary, we have described methods by which a non-equilibrium
nuclear polarisation can be generated and detected in 
a QPC device. Measurements of the nuclear spin relaxation rate
$T_1^{-1}$ are sensitive to the electronic system in the point contact
region, and will show distinctive signatures
of electron-electron interactions that can be used to
distinguish between different proposed scenarios for the 0.7 effect.
Our study shows how NMR methods can be
used to explore novel electronic phenomena in nanoscale semiconductor devices.

\noindent
NRC acknowledges the support of EPSRC grant
GR/S61263/01, and VT the support of TIFR and a DST Ramanujan
Fellowship (sanction no. 100/IFD/154/2007-8).

%\bibliography{./references}

\begin{thebibliography}{10}
\bibitem{footnote2}
We take $I= 3/2$, $n_{\rm nuc} = 4.5\times 10^{22} \mbox{cm}^{-3}$, and the
  (average) hyperfine coupling $A_{\rm s} = 3.3\times 10^{-52} \mbox{J
  m}^3$\cite{paget}. A typical GaAs QPC\cite{abigraham} has $\hbar\omega_y\sim
  2\mbox{meV}$, $\hbar\omega_z\sim 15\mbox{meV}$, such that
  $\sqrt{\omega_y\omega_z}\sim 15\mbox{nm}$.
\bibitem{footnote3}
In general, $T_1$ will depend on the nuclear polarisation itself, through the
  Overhauser shift (\ref{eq:zeeman}). For simplicity, we consider the (change
  in) nuclear polarisation to be small and $Z_{\rm e}$ to be constant.
\bibitem{footnote4}
Nuclei within $w_x$ of the impurity spin are directly coupled to it. Those at
  a distance $R \gtrsim w_x$ remain coupled by RKKY interactions. We find that
  RKKY coupling to the fluctuating impurity dominates the nuclear relaxation
  rate for $R \lesssim \epsilon_F/(k_B T_{\rm K} k_F)$ when $T\lesssim T_{\rm
  K}$, and for $R \lesssim \epsilon_F/[k_BT k_F]$ when $T\gtrsim T_{\rm K}$.
\bibitem{footnote5}
This result is valid for $Z_e \ll k_B T [J_{\rm K}\rho(\epsilon_F)]^2$. In
  the opposite limit, we find $ T_{1}^{-1} = {\pi(k_{B}T)A_{\rm
  d}^{2}S^{2}[J_{\rm K}\rho(\epsilon_{F})]^{2}}/{(\hbar Z_{\rm e}^{2})}$. 
At
  temperatures high compared to the impurity binding energy, $T_1^{-1}$ will
  increase with $T$ and approach that of a non-interacting gas.
\end{thebibliography}
%\bibliographystyle{prsty}

\end{document}